\documentclass[twocolumn,preprintnumbers]{revtex4}

\def \and{\& }

\def\by#1{#1}
\def\and{and }
\def\yr#1{{\bf #1},}
\def\paper#1{#1}
\def\jour#1{{\it #1}}

\def\vol#1{{#1},}
\def\issue#1{}
\def\pages#1{\hbox{#1.}}

\def\JCIS     {J.~Colloid Interface Sci.\ }

\usepackage{amsmath}
\usepackage{amsfonts}
\usepackage{amssymb}
\usepackage{graphicx}
\usepackage{appendix}

\begin{document}

\title{\large Relaxation of surface tension in the liquid-solid interfaces of Lennard-Jones liquids}
\author{\large A.V. Lukyanov, A.E. Likhtman}
\address{School of Mathematical and Physical Sciences, University of Reading, Reading RG6 6AX, UK}

\begin{abstract}
We have established the surface tension relaxation time in the liquid-solid interfaces of Lennard-Jones (LJ) liquids by means of direct measurements in molecular dynamics (MD) simulations. The main result is that the relaxation time is found to be weakly dependent on the molecular structures used in our study and lies in such a range that in slow hydrodynamic motion the interfaces are expected to be at equilibrium. The implications of our results for the modelling of dynamic wetting processes and interpretation of dynamic contact angle data are discussed.
\end{abstract}

\maketitle

The wetting of solid materials by a liquid is at the heart of many industrial processes and natural phenomena. 
The main difficulty in theoretical description and modelling of wetting processes is the formulation of boundary conditions at the moving contact line \cite{Shikhmurzaev2007, Shikhmurzaev1997, Blake2006}. For example, the standard no-slip boundary condition of classical hydrodynamics had to be relaxed to eliminate the well-known non-integrable stress singularity at the contact line \cite{Shikhmurzaev2007, Shikhmurzaev1997, Blake2006, Dussan1976, Hocking1977}. 

The principal parameter of the theoretical description is the dynamic contact angle, which is one of the boundary conditions to determine the shape of the free surface \cite{Shikhmurzaev2007, Shikhmurzaev1997, Blake2006}. The notion of the contact angle has two meanings in macroscopic modelling. One is apparent contact angle $\theta_a$, which is observed experimentally at some distance from the contact line defined by the resolution of experimental techniques (usually about a few $\mu\mbox{m}$) and another one is true contact angle $\theta$ right at the contact line. When the contact line is moving, the apparent contact angle deviates from its static values and becomes a function of velocity. For example, quite often the contact-angle-velocity dependence $\theta_a(U)$ observed in experiments can be accurately described by 
\begin{equation}
\cos\theta_a=\cos\theta_0-a_1\sinh^{-1}(a_2U),
\label{ContactAngle}
\end{equation}
where $a_1, a_2$ are material parameters depending on temperature and properties of the liquid-solid combination, $U$ is the contact-line velocity and $\theta_0$ is the static contact angle \cite{Blake2006}. However useful relationship (\ref{ContactAngle}) may be, it is neither general, due to the well known effects of non-locality \cite{Blake1994, Blake1999}, nor it can be directly used in macroscopic modelling since it is the true contact angle which enters the boundary conditions used in macroscopic analysis. While the apparent contact angle can be experimentally observed, the true contact angle can be only inferred from theoretical considerations or from microscopic modelling such as MD simulations. This is the one of the  main fundamental problems of wetting hydrodynamics, and that problem, despite decades of research, is still far from a complete understanding. The main question still remains open and debates continue: how (and why) does the true dynamic contact angle change with the contact-line velocity? 

The simple hypothesis that $\theta=\theta_0$ has been used in the so-called hydrodynamic theories, for example \cite{Cox1986}, where the experimentally observed changes in the apparent contact angle were attributed to viscous bending of the free surface in a mesoscopic region near the contact line. Some early observations of the meniscus shapes at the contact line have indicated that indeed the meniscus curvature may strongly increase at the contact line \cite{Zanden1994}. The subsequent analysis has shown that while the dynamic contact angle effect may be purely apparent in some cases, it was difficult to rule out variations in the true contact angle. Later on, a numerical study of slip models has shown that whereas viscous bending can contribute to the observed changes in the apparent contact angle, this effect alone is insufficient to explain observations \cite{Wilson2006}. Moreover, recent MD simulations of spreading of LJ liquid drops have shown that the true contact angle does change with the velocity and produce a contact-line-velocity dependence similar to (\ref{ContactAngle}), \cite{Blake2008, Blake2009, Nakamura2013}. The results of MD simulations have been successfully compared against the molecular-kinetic theory (MKT) \cite{Blake2008, Blake2009}. In the MKT, which is also in a good agreement with experiments \cite{Blake2006}, the true contact angle is a function of velocity. This velocity dependence comes from the difference in the probability (asymmetry) of molecular displacements parallel to the solid substrate at the moving contact line, according to the phenomenological assumptions made in the model. The asymmetry is proportional to the contact-line velocity and, on the other hand, to the work done by a macroscopic out-of-balance surface tension force $f_c=\gamma_{LV}(\cos\theta_0 -\cos\theta)$ acting on the contact line. The net result is (\ref{ContactAngle}) with $a_1=2k_B T/\gamma_{LV} \lambda^2$, $a_2=(2k^0\lambda)^{-1}$, where $\gamma_{LV}$ is surface tension at the liquid-gas interface, $k_B$ is the Boltzmann constant, $T$ is the temperature and $k^0$ is the frequency of displacements over the distance $\lambda$, which is regarded as the inverse relaxation time of the surface phase $(k^0)^{-1}=\tau_{LS}$ which is supposed to be proportional to the viscosity $\tau_{LS}\propto\mu$. The key feature of the model is the concentrated force $f_c$ acting on the contact line, similar to the resistive force introduced in \cite{DeGennes1985}, so that the MKT is local. 

\begin{figure}
\begin{center}
\includegraphics[trim=1cm 1cm 1cm 1cm,width=0.88\columnwidth]{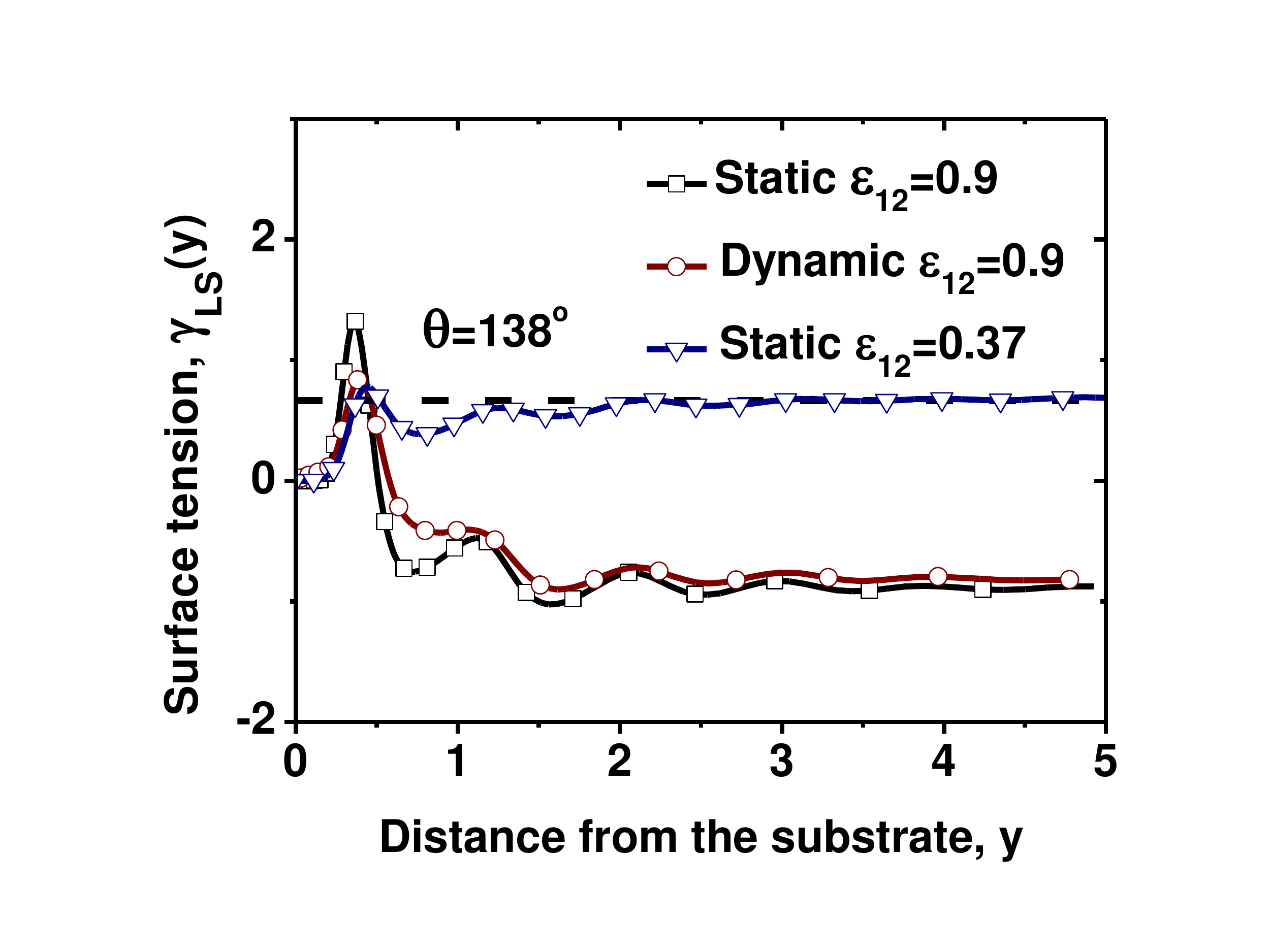}
\end{center}
\caption{Profiles of the integrand of (\ref{stls}) in static (at $\epsilon_{12}=0.9$, $\gamma_{LS}(5)=-0.89$, $\theta_0=15^{\circ}$, and $\epsilon_{12}=0.37$, $\gamma_{LS}(5)=0.68$, $\theta_0=138^{\circ}$,  in the plain geometry) and dynamic (slug geometry, Fig. \ref{Fig3}, averaged over $\Delta z_a=10$ at the contact line, $U=0.1$ and $\epsilon_{12}=0.9$) conditions for a liquid with $N_B=5$ at $T=0.8$. The dashed line shows surface tension level at $\theta_0=138^{\circ}$.} 
\label{Fig0}
\end{figure}

Since the MKT is local, it would be difficult to explain effects of non-locality solely within the model. A more general and potentially universal approach to modelling the dynamic wetting, the interface formation theory, has been proposed by Shikhmurzaev \cite{Shikhmurzaev2007, Shikhmurzaev1997}. The self-consistent macroscopic approach naturally introduces dynamic contact angle through dynamic values of surface tension on forming liquid-solid interfaces. The approach is very appealing and has shown excellent agreement with experimental observations \cite{Blake2002}, but requires the knowledge of macroscopic surface tension relaxation time $\tau_{LS}$ of the liquid-solid interface which is also supposed to be proportional to viscosity, $\tau_{LS} \simeq 4\, \mu \times 10^{-6}\,\mbox{Pa}^{-1}$, \cite{Blake2002}. As a consequence, the theory is truly non-local, that is able to explain effects of non-locality \cite{Blake1994, Blake1999, Lukyanov2006, Lukyanov2007}, with the key characteristic feature, the relaxation tail of the dynamic surface tension with the length scale $\sim U\tau_{LS} (CaSc)^{-1}$, $Ca=\mu U/\gamma_{LV}$, $Sc\simeq 5 (\tau_{LS}/6\times 10^{-9}\,\mbox{s})^{1/2} (1.5\times 10^{-3}\mbox{Pa s}/\mu)^{1/2}$ is a non-dimensional material parameter defining the strength of the interface formation effect $\cos\theta_0-\cos\theta\sim Ca\, Sc$ \cite{Blake2002}.  

In summary, we have at least two principally different models of dynamic wetting, both of them seem to be in a very good agreement with experimental data, \cite{Blake2006, Blake2002}. But, which mechanism does actually determine the dynamic contact angle? To what extent the dynamic interfaces can be in non-equilibrium conditions and contribute into the dynamic contact angle effect?

The key to answering those questions appears to be the surface tension relaxation time $\tau_{LS}$ of the liquid-solid interface, and in this Letter, we directly establish this fundamental parameter by MD simulations. The simulations have been conducted in a model system consisting of LJ particles and/or chain molecules. We investigate $\tau_{LS}$ dependence on liquid viscosity and temperature, and conduct direct MD experiments with dynamic contact angle to get insights into the mechanism of dynamic wetting.    

The MD model we use is similar to \cite{Grest1990} but with the LJ potentials $\Phi_{LJ}^{ij}(r)=4\epsilon_{ij} \left(\left(\frac{\sigma_{ij}}{r}\right)^{12} - \left(\frac{\sigma_{ij}}{r}\right)^{6}  \right)$ and the cut off distance $2.5\, \sigma_{ij}$. Here $i$ and $j$ are either $1$ or $2$ to distinguish between liquid and solid wall particles with the masses $m_i$. Note, hereafter, all units are non-dimensional, the length is measured in $\sigma_{11}$, energy and temperature in $\epsilon_{11}$, mass in $m_1$ and time in $\sigma_{11}\sqrt{ m_1/\epsilon_{11}}$. The beads interacting via LJ potentials are connected into linear chains of $N_B$ beads by the finitely extensible non-linear elastic (FENE) springs, and the strength of the springs is adjusted so that the chains cannot cross each other, $\Phi_{FENE}(x)=-\frac{k}{2}R_0^2\ln\left( 1-\left(\frac{x}{R_0} \right)^2\right)$. Here $R_0=1.5$ is the spring maximum extension and $k=30$ is the spring constant.

\begin{figure}
\begin{center}
\includegraphics[trim=1cm 1cm 1cm 1cm,width=0.87\columnwidth]{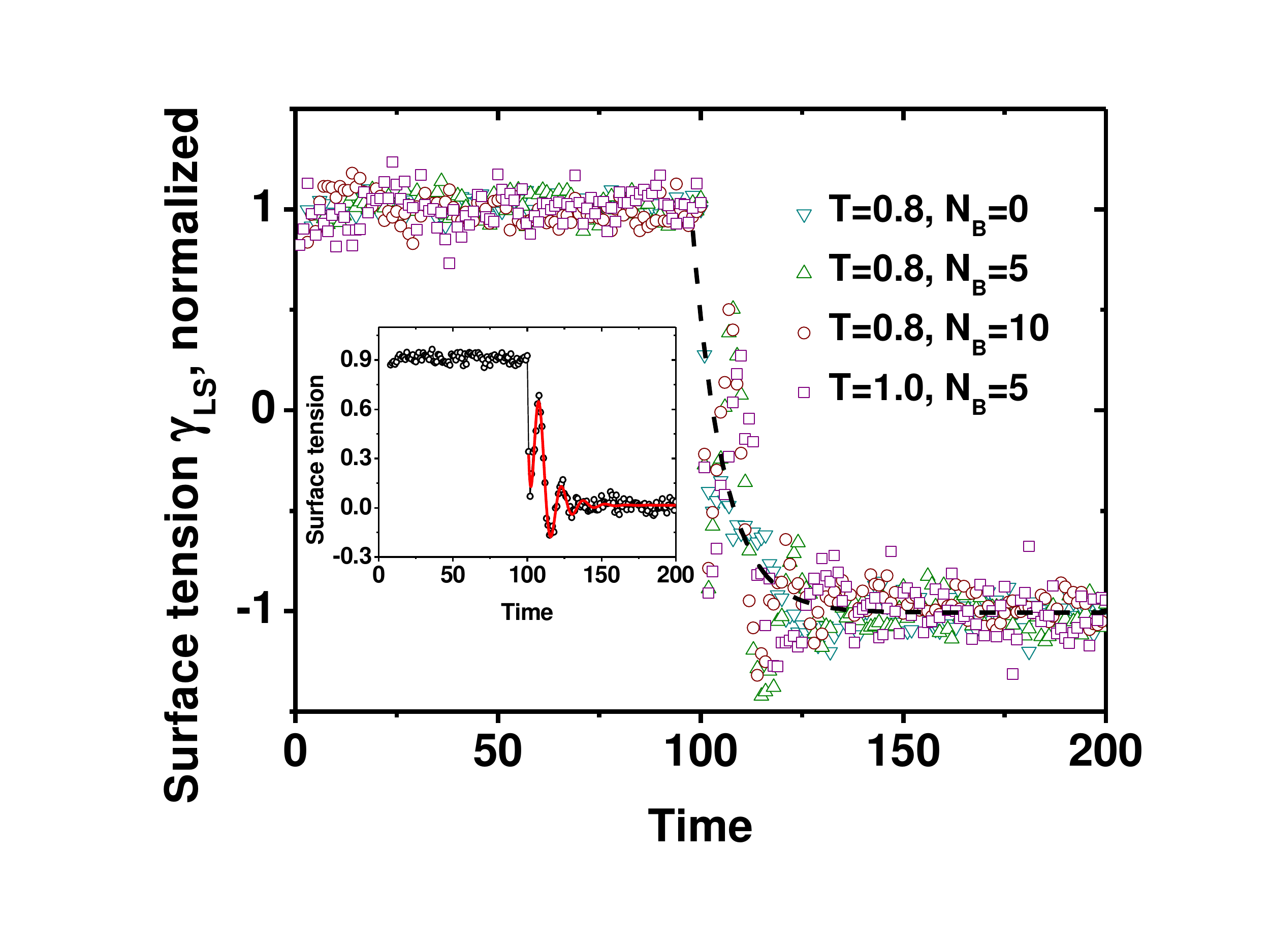}
\end{center}
\caption{Relaxation of liquid-solid surface tension (integrated to $y_m=5$) at different temperatures $T$ and molecular compositions $N_B$ after switching the interaction parameter $\epsilon_{12}$ from $0.2$ to $0.65$ at $t_0=100$ during $\Delta t_s=1$. The data are averaged over $\Delta t_a=1$ and $200$ independent experiments. The dashed line is fit $f_0$. The inset shows individual dependence at $T=0.8$ and $N_B=5$. The solid (red) line in the inset is fit $f_1$.} 
\label{Fig1}
\end{figure}
 
The idea of our MD experiment is simple and is similar to experimentally designed reversibly switching surfaces \cite{Lahann2003}.  First, we equilibrate a square ($L_x=30 \times  L_z=20$) of a liquid film of thickness $L_y \simeq 20$ (the $y$-axis is perpendicular to the film surface and periodic boundary conditions are applied in the $x,z$-directions) consisting of $12000$ particles during $\Delta t_{eq}=5000$ with the time integration step $\Delta t_{MD}=0.01$, which is used in the study. The temperature $0.8\le T\le 1.2$ is controlled by means of a DPD thermostat with friction $\varsigma_{dpd}=0.5$ to preserve liquid motion. The film was positioned between two solid substrates consisting of three $[0,0,1]$ fcc lattice layers of LJ atoms with the shortest distance between the beads $\sigma_{22}$, $\sigma_{22}=0.7$, $m_2=10$ and $\epsilon_{22}=0$. The pressure in the system was kept close to the vapour pressure at given temperature $T$ by adjusting $L_y$ accordingly and making the second wall potential at $y=L_y$ purely repulsive. This has allowed for a small gap between the wall and the liquid phase to establish the gas phase. The solid wall particles were attached to anchor points via harmonic potential $\Phi_a=\xi x^2$, with the strength $\xi=800$ chosen such that the root-mean-square displacement of the wall atoms was small enough to satisfy the Lindemann criterion for melting $\sqrt{<\delta r^2>}<0.15\,\sigma_{22}$  \cite{Barrat2003}. The anchor points in the layer of the solid wall facing the liquid molecules have been randomised in the vertical $y$ direction to increase/vary the surface roughness. The amplitude $\sqrt{<\delta y^2>}=0.1\,\sigma_{22}$ was shown to be sufficient to prevent the substrate from having large and shear-rate divergent/dependent actual slip length \cite{Priezjev2007}. The slip length measured in our experiments, as in \cite{Priezjev2007}, was  $l_{slip}\simeq 2-4\,\sigma_{11}$. After equilibration, parameter $\epsilon_{12}$ of the wall at $y=0$ is changed from one value to another and we observe relaxation of interfacial parameters, including the surface tension.

\begin{figure}
\begin{center}
\includegraphics[trim=1cm 1cm 1cm 1cm,width=0.8\columnwidth]{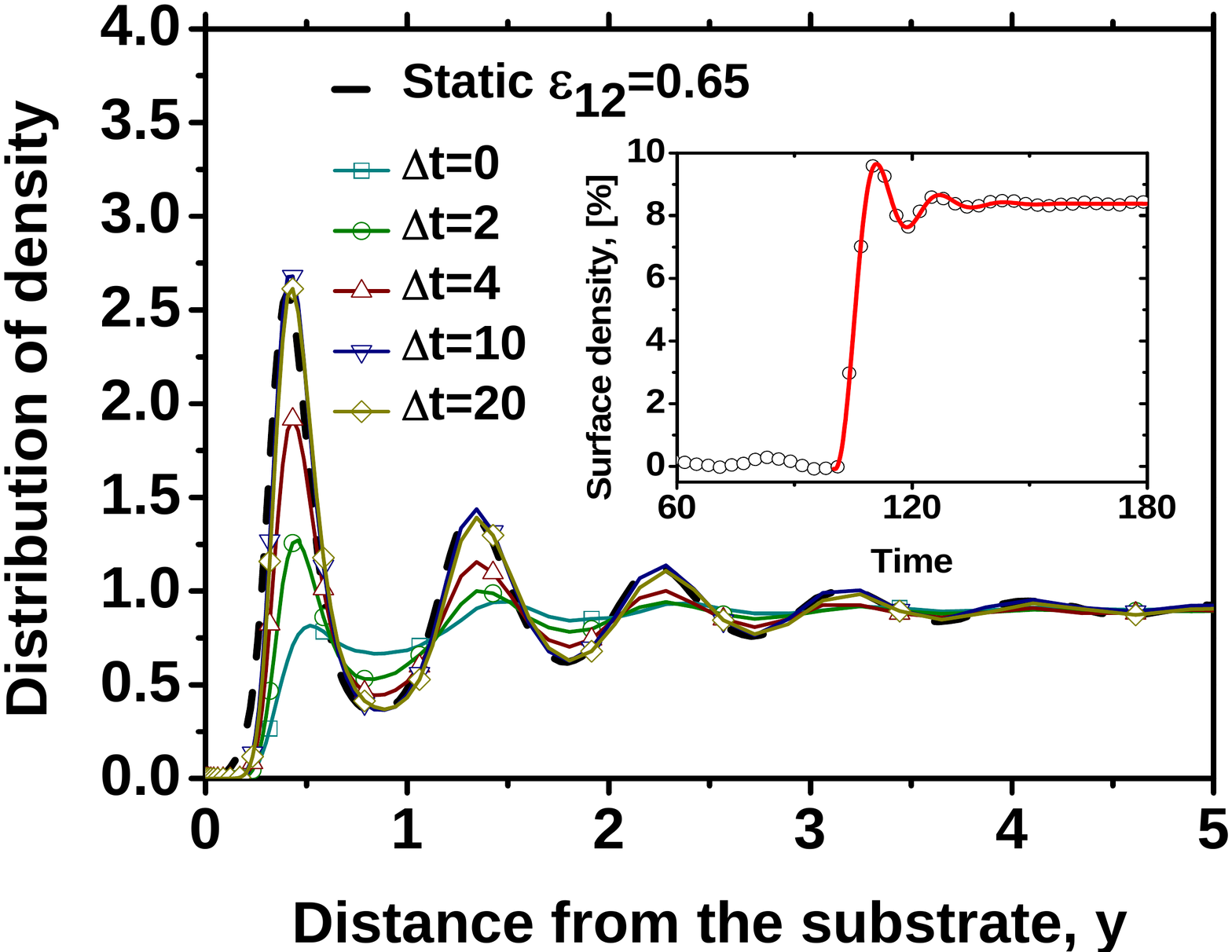}
\end{center}
\caption{Evolution of the density distributions $\rho(y,t)$, the main plot, and surface density $(\rho^{s}(t)-\rho^s(0))/\rho^{s}(0)$, the inset, at $T=0.8$, $N_B=5$ after switching the interaction parameter $\epsilon_{12}$ from $0.2$ to $0.65$ at $t_0=100$ during $\Delta t_s=1$, $\Delta t=t-t_0$. The solid (red) line in the inset is fit $f_1$.} 
\label{Fig2}
\end{figure}

The surface tension of a plane liquid-solid interface is calculated according to \cite{Navascues1979}, in the assumption of the rigid solid substrate  
\begin{equation}
\label{stls}
\gamma_{LS}=\lim_{y_m\to\infty} \int_0^{y_m} \left\{ T_t-T_n - y \rho(y) \frac{d\psi}{dy}\right\} dy.
\end{equation}
Here $\rho(y)$ is distribution of density, $\psi(y)$  is the substrate potential generated by the solid wall particles, $T_{t,n}(y)$ are the tangential and normal components of the microscopic stress tensor evaluated according to \cite{Henderson1982}, all quantities are averaged in the $(x,z)$ plane. We note here that (\ref{stls}) is an approximation in our case of the weakly rough wall consisting of moving particles,  \cite{Navascues1979, Nijmeijer1988, Nijmeijer1990}. So that the numerical procedure has been verified using the Young-Dupree equation by placing a substantially large cylindrical liquid drop (about $30 000$ particles) on the solid substrate and measuring the static contact angle $\theta_0$ applying a three-parameter circular fit $(y-y_0)^2+(z-z_0)^2=R^2$ to the free surface profile. The free-surface profile was defined in the study as the locus of equimolar points. The obtained values of $\theta_0$ were found to be within $3^{\circ}$ of the contact angles calculated directly from the Young-Dupree equation using independently evaluated values of the surface tensions. The liquid-gas $\gamma_{LV}$ surface tension has been calculated using large liquid drops (radius $\sim 30$), similar to \cite{Lukyanov2013}. Typical dependencies of the integrand of (\ref{stls}), $\gamma_{LS}(y)$, in static conditions are shown in Fig. \ref{Fig0} at different values of $\epsilon_{12}$.
  
In the dynamic experiments parameter $\epsilon_{12}$ was switched from $0.2$ to $0.65$ (equivalently $\theta_0=165^{\circ}$ to $90^{\circ}$ at $T=0.8$, $N_B=5$) during $\Delta t_s=1$ with fixed $\sigma_{12}=0.7$. The evolution of surface tension, density distribution and surface density $\rho^s=(\rho_B y_m)^{-1}\int_0^{y_m=5}\rho(y)dy$ are shown in Figs. \ref{Fig1} and \ref{Fig2} for different liquid compositions and temperatures. One can see that in general the relaxation is very quick and almost independent of viscosity of the liquid at the first glance (the results are insensitive to lowering  $\varsigma_{dpd}$ to $0.3$). Simple fit, $f_0=C_1+C_2\exp(-(t-t_0)/\tau_0)$, applied on average to normalised surface tension evolution data reveals $\tau_0=7.8$. The individual dependencies, inset Fig. \ref{Fig1}, reveal more complex behaviour, which can be approximated by $f_1=C_1+C_2\exp(-(t-t_0)/\tau_1)+C_3\exp(-(t-t_0)/\tau_2)\sin(\omega(t-t_0)+\phi_0)$, Table \ref{Table1}. One can see that both  $\tau_1$ and $\tau_2$ are almost independent of molecular structure/viscosity despite the seventy-fold variation in $\mu$. The observed values of $\tau_1$ (the major relaxation) are close to the relaxation times found in the free surfaces of LJ liquids, \cite{Lukyanov2013}, and thus correspond to the local relaxation on the length scale of the individual density peaks, Fig. \ref{Fig2}, that is on the beads level rather than on the level of the whole molecules. This is similar to the multi-scale relaxation commonly observed in polymer dynamics, \cite{Likhtman2007}. In our case, initial, early times relaxation is defined by the mean square displacement of individual monomers over relatively short distance of the order of the half of the distance between the density peaks ($\Delta y=0.5$), Table \ref{Table1}, while the liquid viscosity is defined by the much slower molecular relaxation. This is also consistent with the weak dependence on the destination value $\epsilon_{12}(t>t_0)$, Table \ref{Table1} the last row. 

The second, oscillatory relaxation, $\tau_2$ and $\omega$, is likely to be due to the collective excitation of the particle motion triggered by the sharp change of the solid wall potential, since the amplitude of oscillations decreases with increasing the switching time interval $\Delta t_s$. In this case, $\tau_2$ is simply the time during which the excited wave of frequency $\omega$ travels some distance $l_2$ comparable to the interfacial layer width. Indeed, the product $\tau_2\omega=2\pi l_2/\lambda_2$, where $\lambda_2$ is the wave length, varies within $3.1\le\tau_2\omega\le5.8$, $<\tau_2\omega>=4.4$, Table \ref{Table1}. Then on average $<l_2/\lambda_2>\simeq0.7$ which means that the wave length of the excitations is roughly the width of the interfacial layer. 

We would like to note that the observed weak dependence $\tau_{1,2}(\mu)$ is in contrast to the relaxation time scaling $\tau_{LS}\propto \mu$ found in the MKT and in the interface formation theory. While the first peak density characteristic time scale found in the MD simulations \cite{Blake2009} $\tau_{dp}\simeq 16.5$ is roughly comparable to our results, Table \ref{Table1}, the observed weak dependence $\tau_{1,2}(\mu)$ rules out possible connections between $\tau_{LS}$ and the MKT parameter $(k^0)^{-1}$.

\begin{figure}
\begin{center}
\includegraphics[trim=1cm 1cm 1cm 1cm,width=0.8\columnwidth]{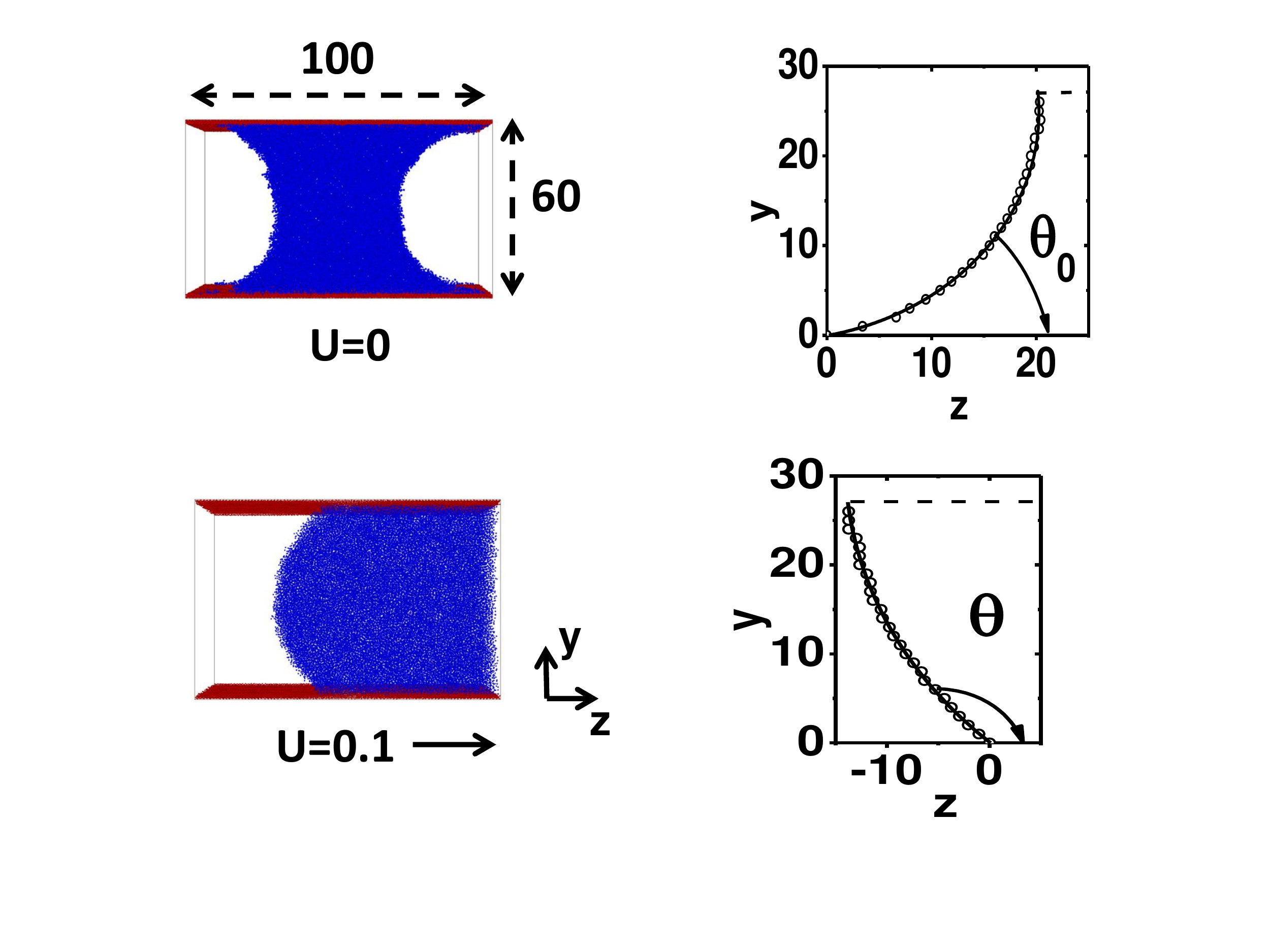}
\end{center}
\caption{Snapshots and free surface profiles (the circular fits) in static and dynamic ($U=0.1$) situations at $T=0.8$, $N_B=5$ and $\epsilon_{12}=0.9$ ($\theta_0=15^{\circ}$). The observed static and dynamic contact angles are $\theta_0=12\pm3^{\circ}$ and $\theta=138\pm4^{\circ}$. The direction of the moving solid wall particles is indicated by the arrow.} 
\label{Fig3}
\end{figure}

\begin{table*}[t]
\begin{tabular}{ | c | c | c | c | c | c | c | c | c | c | c | c | c  | c | c |}
\hline
  $T$   & $N_B$  & $\gamma_{LV}$    & $\rho_B$  & $\mu^{\dag}$ &  $\tau_1$   &  $\tau_2$   &  $\omega$ &  $\tau_1^{\rho}$  &  $\tau_2^{\rho}$ &  $\omega^{\rho}$ & $\sqrt{<r_M^2>}$\\  
\hline
  $0.8$ & $1^{\dag\dag}$    & $0.36\pm0.02$ & $0.73$  & $1.2$ &  $2.4\pm0.6$  &  $15.3\pm2.2$     &  $0.20$ &  $5.5\pm 0.5$  &  $16.1\pm0.4$   &  $0.20$  & $0.61$ \\ 
\hline
  $0.8$ & $5$    & $0.92\pm0.04$ & $0.91$  & $10.5$  &  $3.9\pm0.3$  &  $10.7\pm0.8$      &  $0.43$ &  $3.8\pm0.1$  &  $9.3\pm0.2$   &  $0.39$  & $0.40$ \\ 
\hline
  $0.8$ & $10$   & $1.01\pm0.05$  & $0.93$  & $20.2$  &  $4.3\pm0.3$  &  $9.2\pm0.7$   &  $0.48$ &  $3.6\pm0.1$  &  $8.7\pm0.2$   &  $0.44$  &  $0.42$ \\ 
\hline
  $0.8$ & $15$   & $1.05\pm0.05$  & $0.93$  & $30.1$  &  $3.5\pm0.6$  &  $13.8\pm1.8$   &  $0.35$ &  $4.6\pm0.3$  &  $12.1\pm0.5$   &  $0.32$  &  $0.38$ \\ 
\hline
  $0.8$ & $20$   & $1.08\pm0.05$  & $0.94$  & $41.1$  &  $3.9\pm0.5$  &  $9.2\pm1.1$   &  $0.45$ &  $3.8\pm0.2$  &  $9.2\pm0.3$   &  $0.40$  &  $0.40$ \\ 
\hline
  $0.8$ & $30$   & $1.1\pm0.06$  & $0.94$  & $68.2$  &  $3.6\pm0.2$  &  $5.3\pm0.3$   &  $0.79$ &  $3.0\pm0.1$  &  $5.5\pm0.2$   &  $0.72$  &  $0.38$ \\ 
\hline
  $1.0$ & $5$    & $0.71\pm0.03$ & $0.86$  & $5.7$   &  $4.0\pm0.5$  &  $12.2\pm1.2$       &  $0.33$ &  $4.5\pm0.2$  &  $13.2\pm0.3$   &  $0.3$  &  $0.56$ \\ 
\hline
  $1.0$ & $8$    & $0.78\pm0.03$ & $0.87$  & $9.4$   &  $3.7\pm0.5$  &  $13.2\pm1.2$       &  $0.33$ &  $4.0\pm0.2$  &  $11.9\pm0.3$  &  $0.32$ &  $0.54$\\ 
\hline 
  $1.0$ & $50$    & $0.92\pm0.05$ & $0.89$  & $61.8$   &  $4.4\pm0.8$  &  $27.5\pm2.9$       &  $0.21$ &  $5.1\pm0.3$  &  $27.2\pm1.2$  &  $0.21$ &  $0.55$\\ 
\hline     
  $1.2$ & $5$    & $0.52\pm0.03$ & $0.79$  & $3.8$   &  $2.3\pm0.6$  &  $20.7\pm3.2$       &  $0.23$ &  $4.5\pm0.3$  &  $17\pm0.7$   &  $0.24$  &  $0.56$  \\ 
\hline 
  $0.8$ & $5^{\dag\dag\dag}$    & $0.92\pm0.04$ & $0.91$  & $10.5$  &  $4.3\pm0.3$  &  $9.6\pm0.8$      &  $0.44$ &  $3.7\pm0.1$  &  $8.3\pm0.2$   &  $0.42$  & $0.44$ \\ 
\hline                  
\end{tabular}
\caption{Parameters of the liquids (equilibrium surface tension $\gamma_{LV}$, bulk density $\rho_B$, dynamic viscosity $\mu^{\dag}$) and characteristic times of the liquid-solid interface formation ($\tau_{1,2},\,\omega$ for the surface tension and $\tau_{1,2}^{\rho},\,\omega^{\rho}$ for the surface density $\rho^{s}$ applying fit $f_1$) at different molecular compositions (number of beads $N_B$) and temperatures $T$. $^{\dag}$ Viscosity was obtained as in \cite{Likhtman2007} at the bulk conditions. The last column is the end-monomer mean-square displacement $\sqrt{<r_M^2>}$ during $\Delta t=\tau_1$ across the interface at the bulk conditions. $^{\dag\dag}$ $400$ independent experiments. $^{\dag\dag\dag}$ $\epsilon_{12}(t>t_0)=0.9$.}
\vspace{12pt}
\label{Table1}
\end{table*}

\begin{figure}
\begin{center}
\includegraphics[trim=1cm 1.5cm 1cm 2.5cm,width=0.9\columnwidth]{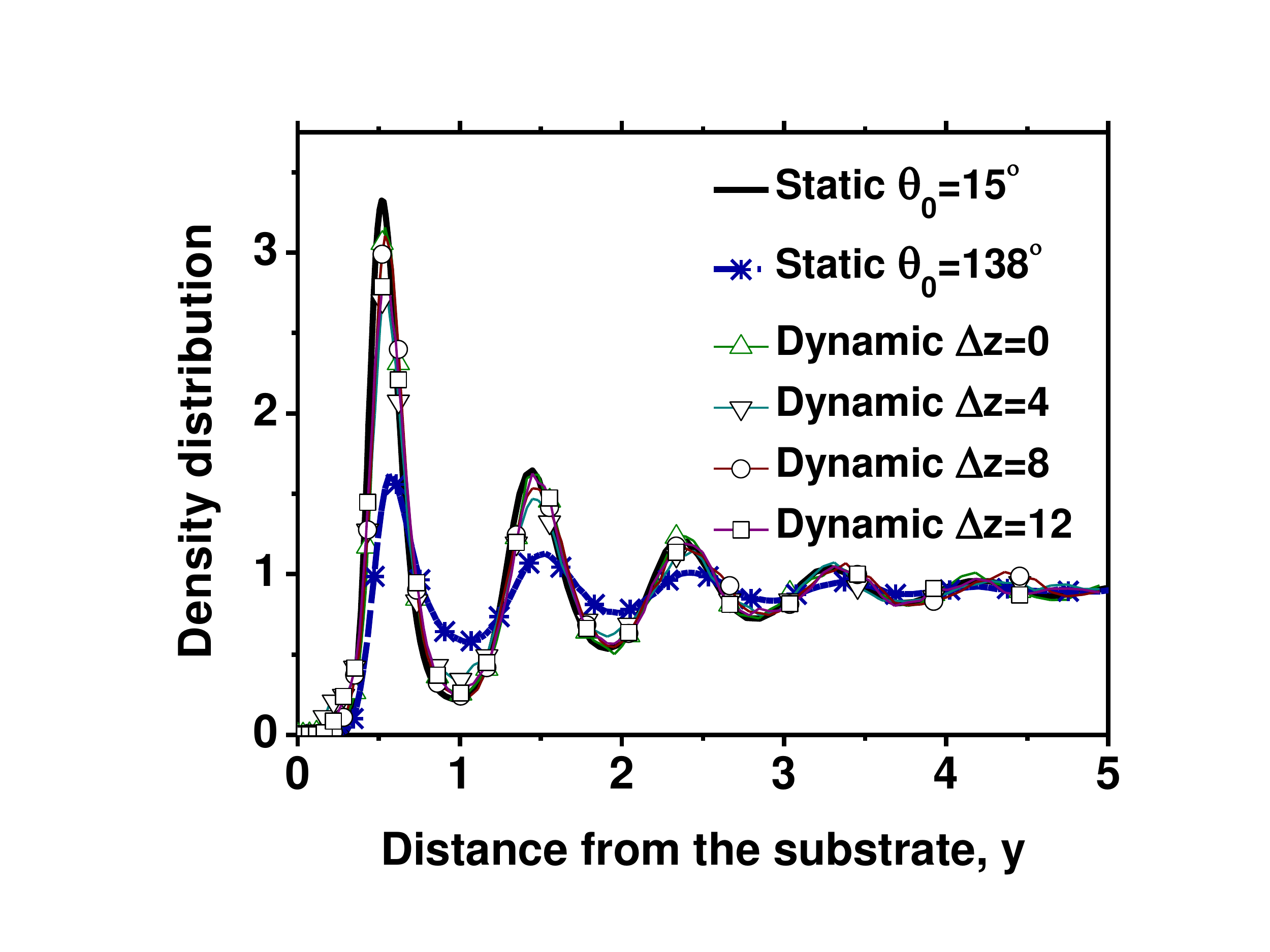}
\end{center}
\caption{Density distribution at different distances from the contact line $\Delta z$ averaged over $\Delta z_a=1.25$ in the dynamic case shown in Fig. \ref{Fig3} at $U=0.1$, $\epsilon_{12}=0.9$ (slug geometry) and in static conditions (plain geometry) at $T=0.8$, $N_B=5$, and $\epsilon_{12}=0.9$ ($\theta_0=15^{\circ}$) and $\epsilon_{12}=0.37$ ($\theta_0=138^{\circ}$).} 
\label{Fig4}
\end{figure}

The relaxation times revealed by the dynamic experiments directly imply that in the liquid compositions used in our study, in the slow hydrodynamic motion, parameter $U\tau_{LS}/L << 1$ ($L>>1$ is any macroscopic length scale) and surface tension is expected to be at equilibrium. This in turn implies that the dynamic surface tension is unlikely to be the cause of dynamic angle in our case, $Sc<<1$. To verify this conclusion, we have performed a series of MD experiments with a large cylindrical liquid slug ($60000$ particles) forced between two identical rough solid plates, Fig. \ref{Fig3}. The geometry is periodic in the $x$-direction with reflective boundary conditions at the simulation box ends in the $z$-direction. The solid wall particles are moving with velocity $U$ in the $z$-direction to mimic forced wetting regime. After initial equilibration during $\Delta t_{eq}=5000$, we measure the dynamic contact angle and interface parameters in steady conditions. The dynamic contact angle can be clearly seen in the snapshot and in the developed free surface profile, Fig. \ref{Fig3}. This is an extreme case ($Ca=1.1$) of typical profiles observed in the case of long-chain molecules when the dynamic contact angle $\theta$ is changing monotonically with the substrate velocity $U$ from its equilibrium value. We have checked that the system size has already no dramatic effect on the observed contact angle. For example, in a similar case $\epsilon_{12}=0.65, \theta_0=90^{\circ}, U=0.1$, at $60000$ particles $\theta=143.7\pm3^{\circ}$, at $40000$ particles, $\theta=142.4\pm3^{\circ}$, while at $10000$ particles $\theta=129\pm3^{\circ}$. 

The direct measurements of surface tension and distribution of density, in the case shown in Fig. \ref{Fig3}, right after the contact line (the contact line width is taken at $\Delta_{cl}=6$ counting from the intersection of the free surface and the substrate at $z_{cl}=0$, Fig. \ref{Fig3}, just to fully cover the interfacial zones of both interfaces) are shown in Figs. \ref{Fig0}, \ref{Fig4}. One can see that indeed while the liquid motion has some effect on the first layer of particles, the overall effect is not large, and both the surface tension and the density are close to equilibrium, and far away from the values in the case $\epsilon_{12}=0.37$ ($\theta_0=138^{\circ}$ similar to the observed dynamic angle). How had then that dynamic angle (different from $\theta_0=15^{\circ}$) been generated? We analysed the tangential force acting on the interface molecules in the region ($0\le y\le y_m=2$, $z_{cl}\le z\le z_{cl}+\Delta_{cl}$) at the contact line. We found that the tangential force $f^{cl}_z$ acting on the liquid from the solid substrate is concentrated within $\Delta_{cl}$ and then drops significantly. This is not a coincidence, of course, since $\Delta_{cl}\simeq l_{slip}$. The value of the force per unit length of the contact line is found to be sufficient to generate the observed contact angle according to the balance of all forces acting on the contact line, the modified Young-Dupree equation, that is $f_z^{cl}=1.52\pm 0.13$ and from $\gamma_{LV}\cos(\theta)=-\gamma_{LS}-f_z^{cl}$, $\theta=133^{\circ}\pm 10^{\circ}$. But this effect will need further studies.

In conclusion, we have directly established relaxation time of the liquid-solid interfaces in a model system consisting of LJ molecules. The relaxation time, importantly, appears to depend  very weakly on the molecular structure and viscosity and is found to be in such a range that interfacial tension $\gamma_{LS}$ should be in equilibrium in slow hydrodynamic motion. This has been also verified in the MD experiments on dynamic wetting, where the dynamic contact angle was observed. Our results have direct repercussions on the theoretical interpretation and modelling of the dynamic contact angle. 

{\bf Acknowledgement.} This work is supported by the EPSRC grant EP/H009558/1. The authors are grateful to Y. Shikhmurzaev and T. Blake for useful discussions.

\end{document}